# A NEW TEACHING MODEL FOR THE SUBJECT OF SOFTWARE PROJECT MANAGEMENT

*M. Rizwan Jameel Qureshi[a] *Muhammad Rafiq Asim[b] *Muhammad Nadeem[c] Asif Mehmood[d]
Email: rjamil@ciitlahore.edu.pk, Email:mrasim@ciitlahore.edu.pk, Email: nahmed@ciitlahore.edu.pk,Email:asifmehmood_mcs@yahoo.com
*Department of Computer Science, COMSATS Institute of Information Technology M. A. Jinnah Campus, Defence Road off Raiwind Road, Lahore Pakistan.

*ABSTRACT:* Software (SW) development is a very tough task which requires a skilled project leader for its success. If the project leader is not skilled enough then project may fail. In the real world of SW engineering 65% of the SW projects fail to meet their objectives as in [1]. The main reason is lack of training of the project mangers. This extreme ratio of failure can be reduced by teaching SW project management (SPM) to the future project managers in the practical manner, so that they may be skillful enough to handle the project in a better way. This paper intends to propose a model to be used to teach SPM to the student of SW engineering to reduce the failure rate of projects.

## 1. INTRODUCTION

SPM is an important course that it is very useful in a sense that it provides well mature and much disciplined distraction to manage the SW project. SPM requires a lot of practice instead of cramming theories. Most of the Pakistani Universities focus on teaching the theories to the students without practical exposure. Since such students do not have exposure of the actual problems being faced in SW development, which in turn leads to failure SW projects.

SPM is combination of knowledge and skills. SPM includes many tasks such as planning, tracking and control of the SW development projects [2]. It is very much similar to flying of a plane by an untrained/inexperienced pilot. A pilot is not expected to fly the plane only by means of instructor or the text. Instead it requires that the trainee should be trained in such an environment where he could come across the real time problems faced in flying a plane that is a way computer simulation. These are the compulsory part of the pilots training where pilot gets familiar to the situations which can arise in the real world fight [3].

Research has provided that poor management can increase the overall cost of the SW development more rapidly than other any factor [4]. This development cost could be minimized by training the future project managers in a better way to get solid knowledge and practice of SPM.

It is the need of day that students may be involved in some projects during their studies as a practical training.

The reminder of paper is organized as follows, the next section describes some brief literature review and in section 3, the authors have described the statement of the problem and in section 4, solution towards that problem is given. In section 5, the solution is validated by the means of survey and conclusion is provided in section 6.

## 2. Related works

SPM requires strong method to be used to teach it effectively to the students of SW engineering. Some efforts have been done regarding it but these efforts are not enough and have some limitations. In the literature some research has already been done on the issue of teaching SPM effectively to the students of SW engineering.

In [5], Longjun Huang and others have presented a solution, in which they introduced the idea of "learning by doing". It focuses on practice instead of cramming and it turns passive learning into active learning. It also changes student as a center of the teaching process.

The authors [3] proved that simulations can be used to teach SPM effectively. They proposed "System Dynamics Simulation Training Tool" for the training of project managers. It benefited the student by means of duration, repeatability, realism, extensibility and measurability. Its limitation is clear objectives.

As in [6], Pantelis M. Padasopulos and others have presented "e-Case-SPM Web environment learning Tool" for teaching SPM. There are four major designs

In [7], Eduardas Baresia and other suggested that teaching by designing can benefit both students and faculty as well. It provides latest information and requirements of the SW Industry (SI). The limitations of this solution are the use of waterfall model which has many disadvantages.

The authors, in [8] proposed that project planning can be easily done using extreme programming. Groups are formed and they altogether complete the task of project planning. In this way students are able to excel by the communication and interaction among them.

James McDonals in [9] has shown some differences and similarities between teaching SPM in industrial workshop and in academic environment. According to him, academic version students can have time for extensive reading between their classes, which is not possible in an industrial workshop.

As in [10], the authors illustrated the idea of Interactive Learning Environment. It provides students with the ability to perform the process involved in SPM. It defines three main roles which are Human resource Manager (HRM), Planning and Controlling Manager (PCM), Production & Testing and Quality Assurance Manager (PTQM). Students can use these roles and their responsibilities in the form of game. They perform these responsibilities within time and budget. So using this game their leaning capabilities are improved in a short time.

The authors in [11] proposed that students may be involved in the client-based team projects in their classes. Starting from first week of their semester they are grouped and the projects are assigned to them, whereas the teacher plays the role of upper project management. Students in the groups are assigned different roles to perform. For the project to be completed successfully it requires the combined effort made by students and the class instructor as well. This idea enables students to learn the experience to handle difficulties and real time problems that may arise in the SPM in SI.

As in [12], the authors described the "Practicum in SPM" a course offered in the university to the students of "Masters in SPM". The students are assigned the projects in which undergraduate students work as the team workers while graduate students are project managers. Teachers are involved as client of the projects. In this way student of graduate classes learn by leading the team of undergraduate students. Nevertheless, there may arise some limitations such as no real SW cost estimation. There are a large number of projects in which there is small number of enrolled graduate students so many projects may be left without a project manager.

In [13] the authors proposed that students should use Lego bricks to build some building in collaboration with the instructor. In this way they would be able to learn the concepts of the project management. They will also face some problems that may arise during selecting a landmark and towards its completion.

Table 1 gives a brief description of the literature reviewed regarding this paper including project title and some limitations which are found in them.

Table 1 Comparison of brief literature review

| Title of paper | Limitations |
|---|---|
| Project driven teaching model [5]. | • No interaction with industry. |
| Improving SPM skills by using a SW project simulator [3]. | • Suitable only for planning, tracking and control of project.<br>• Training is less effective if objectives are not clear. |
| Case-based interaction on the web for teaching. [6] | • Much workload for students.<br>• Complexity<br>• Diverse nature of the students. |
| Research and development of teaching SW engineering processes [7]. | • Too much documentation needed.<br>• No industry involvement.<br>• Usage of waterfall model |
| Learning project planning the agile way [8]. | • Scalability.<br>• Accurate estimation. |
| Teaching SPM in Industrial and academic environment [9]. | • No extensive reading of literature by the trainees.<br>• Hard to follow due to faster pace |
| | • No in- depth knowledge.<br>• No hand on practice. |
| Interactive learning environment designing for SPM teaching [10]. | • No industry collaboration.<br>• Game based.<br>• Not evaluated yet. |
| A case study of classroom experience with-client based team projects [11]. | • Complexity<br>• Difficult to mange in classroom environment.<br>• Much time consuming.<br>• Team work effort required |
| Practicum in SPM- An endeavor to effective and pragmatic SPM educations [12]. | • No industry collaboration.<br>• For small team size only.<br>• Difficult cost estimation. |
| Teaching SPM using simulation [13]. | • Construction oriented.<br>• Much time costing.<br>• No industry collaboration |
| University/Industry collaboration in developing a simulation based SPM training course [2]. | • Not case based.<br>• Not project driven. |

## 3. Problem Statement
How to teach SPM effectively with solid hands on practice?

## 4. Student/Industry Interaction Teaching Model
The authors proposed solution is a model for teaching SPM. This model is based on the interaction between students, instructor and the SI. The proposed model creates an atmosphere for the students which will act as an excellent tool for developing an interest of getting SPM and hands-on practice.

### a) Objectives of the model
The main objectives of the proposed model are to:
- ❖ Develop a complete project plan.
- ❖ Provide knowledge of the SPM.
- ❖ Provide hands-on practice.
- ❖ Enable to find out SW cost estimation and its scheduling.
- ❖ Enable to handle real life SW projects.
- ❖ Better resource utilization.
- ❖ Minimize the number of errors.
- ❖ Decrease the overall SW project cost.
- ❖ Develop quality SW.
- ❖ Handel the team in a better and systematic manner.

### b) Components of the model
There are three main stake holders in the model proposed by the authors.
- ❖ *The instructor of the class.* Instructor is the SW project manager and responsible for the monitoring of the SW project.

- *The students.* They are working in groups and changing their roles as project progresses, designer, developer etc. Roles are assigned by the instructor.
- *SI.* Acting as the customer/client of the SW being developed.

**c) Teaching method**

The students are divided into groups. Each group consists of 6 to 8 students. Then there is preliminary interaction session between each group and different SW houses. The projects are taken from the SW houses and SW houses works as a customer of the SW project. After this instructor assigns different roles to the students in a group. There has to be communication between groups and SW houses in terms of requirements gathering sessions. During the development of SW, the instructor deeply examines the progress of the SW development and other activities. Students in a group work according to the roles assigned by their instructor. When requirement gathering is complete, prototype of the SW is developed and evaluated by the customer. Feedback is received by the students and revised version of SW is developed according to the requirements of the customer.

This process continues as shown in figure 1 until the customer is satisfied.

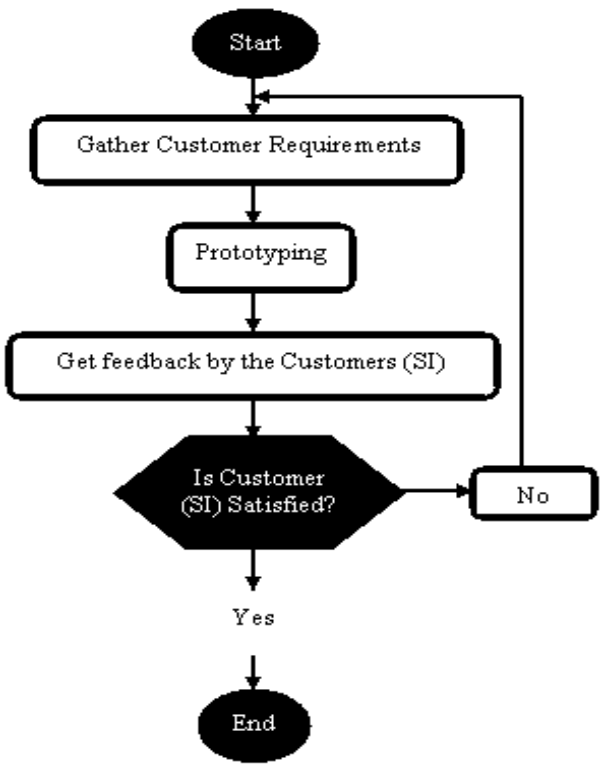

Figure 1 Prototype Model

**d) Benefits of the model**

Benefits in teaching SPM:
- It provides a real life SW projects development environment.
- It provides a good hands-on experience of SW development.
- This model is best suitable for cost estimation of the project.
- In this model students can work in different roles.
- The model enables the student to remain in touch with the updates of SI.
- Students are enabled to get the latest information regarding changes in the SI.
- It is a quick method of teaching SPM practically.

## 5. Validation

Survey was conducted for the validation purpose. A questionnaire consisting of 14 questions was distributed among the different Pakistani institutions. After that results were gathered and analyzed through an analysis tool SPSS for cumulative evaluation. Likert scale is given in the following Table 2.

Table 2 Likert Scale

| 1 | Strongly disagree |
|---|-------------------|
| 2 | Disagree |
| 3 | Neutral |
| 4 | Agree |
| 5 | Strongly agree |

It may be mentioned that during the course of validation likert scale 1 & 2 counted towards opposing and the likert scales 4 & 5 are considered to be supporting the query in question.

Q. # 1: Is it necessary to teach SPM to the students of SW engineering?

Results of question 1 given in Table 3 showing that 60 percent of the people were supportive to question 1 that it is necessary to teach SPM to the students of SW engineering whereas 8 percent of the people were opposed to teach SPM to the students of the SW engineering. The percentage of the people who have neutral opinion is 32 percent.

The conclusion of the survey of this question is that essential to teach SPM to the students of SW engineering.

Following is the Table 3 showing the results obtained for the question 1.

Table 3 Necessity to Teach SPM

| | Likert Scale | Frequency | Percent | Cumulative Percent |
|---|---|---|---|---|
| Valid | 2 | 2 | 8.0 | 8.0 |
| | 3 | 8 | 32.0 | 40.0 |
| | 4 | 11 | 44.0 | 84.0 |
| | 5 | 4 | 16.0 | 100.0 |
| | Total | 25 | 100.0 | |

Following bar chart 1 is showing the results obtained for question 1.

Bar chart 1 Necessary to Teach SPM

Q. # 2: How much ineffective are the existing teaching models of Pakistani Universities, adopted for teaching SPM?
Table 4 is clearly showing 76 percent of the people declared the existing teaching model of Pakistani Universities to be ineffective whereas 12% people told the existing system to be effective. The people who were neutral during survey about this question are 12 percent.

This highly supportive percentage (76 percent) shows that the existing teaching models of SPM in Pakistani Universities are ineffective.

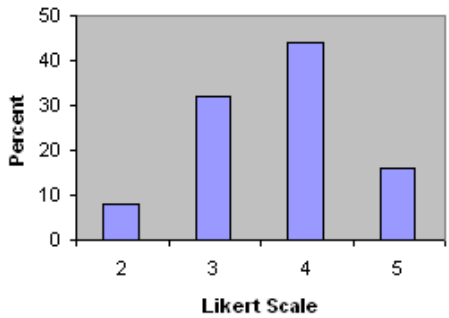

Table 4 Ineffective Teaching Models

| | Likert Scale | Frequency | Percent | Cumulative Percent |
|---|---|---|---|---|
| Valid | 1 | 1 | 4.0 | 4.0 |
| | 2 | 2 | 8.0 | 12.0 |
| | 3 | 3 | 12.0 | 24.0 |
| | 4 | 14 | 56.0 | 80.0 |
| | 5 | 5 | 20.0 | 100.0 |
| | Total | 25 | 100.0 | |

Following bar chart 2 shows the results obtained in question 2.

Bar chart 2 Ineffective Teaching Models

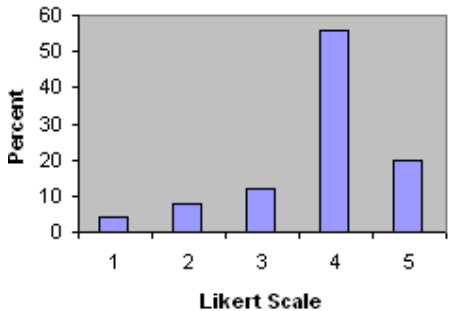

Q. # 3: Do you think that there is need for a new teaching model of SPM?
Results about question 3 are shown in Table 5.
Regarding question 5 the outcome of the survey is given in Table 5 which is obviously showing that there is a need for a new teaching model for SPM. This is because 64 percent of the people were supporting the need of a new teaching model to teach SPM to the students of SW engineering. At the same time 24 percent of the people disagreed to teach SPM by a new method. However, only 12 percent of the people remained on the fence.

The resultant effect of this survey indicates that it is compulsory to teach SPM with a new method.
The results of the Table 5 are also displayed by the Bar chart 3.

Table 5 Need for New Teaching Model of SPM

| | Likert Scale | Frequency | Percent | Cumulative Percent |
|---|---|---|---|---|
| Valid | 2 | 6 | 24.0 | 24.0 |
| | 3 | 3 | 12.0 | 36.0 |
| | 4 | 5 | 20.0 | 56.0 |
| | 5 | 11 | 44.0 | 100.0 |
| | Total | 25 | 100.0 | |

Bar chart 3 Need for New Teaching Model of SPM

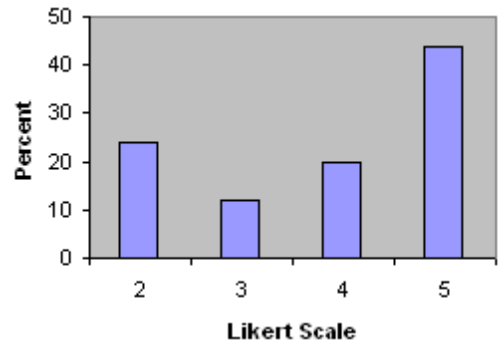

Q. # 4: Does SPM require hands-on practice?
As far as question 4 is concerned, there are 60 percent of the people who informed during survey that SPM could not be taught with out hands-on practice. This is due to the fact that this course will not give practical exposure without the involvement of the students in a practical activity. Even as only 8 percent of the people were against the hands on-practice. The percentage of the people who put their view to oppose this question was 8 percent only whereas 32 percent of the people remained neutral. By comparing and contrasting the supportive and unsupportive percentages, one can easily understand that the subject, SPM, definitely require hands-on practice to learn this course effectively.

Table 6 Hands-on Practice

| | Likert Scale | Frequency | Percent | Cumulative Percent |
|---|---|---|---|---|
| Valid | 2 | 2 | 8.0 | 8.0 |
| | 3 | 8 | 32.0 | 40.0 |
| | 4 | 9 | 36.0 | 76.0 |
| | 5 | 16 | 24.0 | 100.0 |
| | Total | 25 | 100.0 | |

The following bar chart 4 shows the results given in Table 6.
Bar chart 4 Hands-on Practice

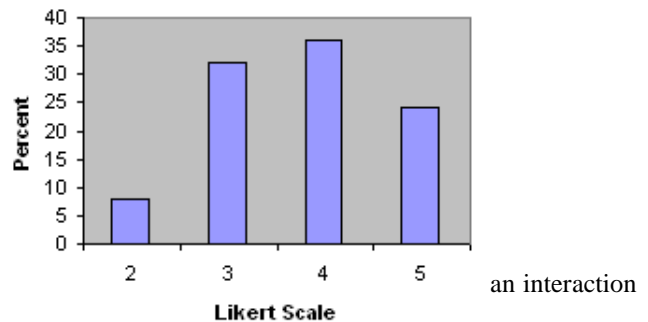

an interaction

Table 7 is the response for the question 5 which visibly gives the picture that 60 percent of the people were of the view that the students should have a communication channel to interact with SI. Concurrently 8 percent of the people disagreed with the interaction of the students of SW engineering to interact with SI.

Since this is a little bit opposing percentage as compared to supporting one, therefore it is concluded that the SW engineering students must have liaison with SI to get the full advantages of studying SPM. Meanwhile there were 32 percent of the people in neutral decision under the query in question.

Table 7 Interaction with SI

| | Likert Scale | Frequency | Percent | Cumulative Percent |
|---|---|---|---|---|
| Valid | 2 | 2 | 8.0 | 8.0 |
| | 3 | 8 | 32.0 | 40.0 |
| | 4 | 11 | 44.0 | 84.0 |
| | 5 | 4 | 16.0 | 100.0 |
| | Total | 25 | 100.0 | |

Bar chart 5 elaborated the results of Table 7 in graphical form.

Bar chart 5 Interaction with SI

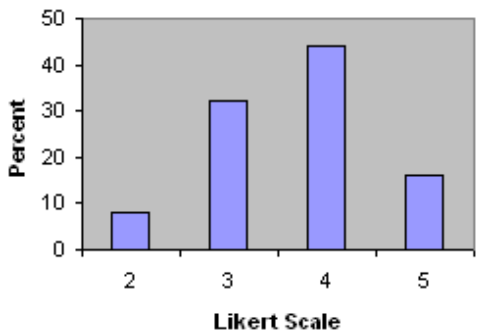

Q. # 6: Is it useful for the students of SPM to take SW project direct from SI?

There is a very interesting conclusion depicting from the Table 8. This table explains that 72 percent of the people put up their answers in favor of the truth that the students must get the SW project from the SI directly. At the same time only 8 percent of the people stated that the students should not get SW project from the SI openly. Only 20 percent of the people who voted this question while showing their-self impartiality.

According to this extremely encouraging percentage (72 percent), it is very easy to judge the importance to get SW project from SI openly.

Table 8 Taking Project from SI

| | Likert Scale | Frequency | Percent | Cumulative Percent |
|---|---|---|---|---|
| Valid | 1 | 1 | 4.0 | 4.0 |
| | 2 | 1 | 4.0 | 8.0 |
| | 3 | 5 | 20.0 | 28.0 |
| | 4 | 13 | 52.0 | 80.0 |
| | 5 | 5 | 20.0 | 100.0 |
| | Total | 25 | 100.0 | |

Bar chart 6 Taking Project from SI

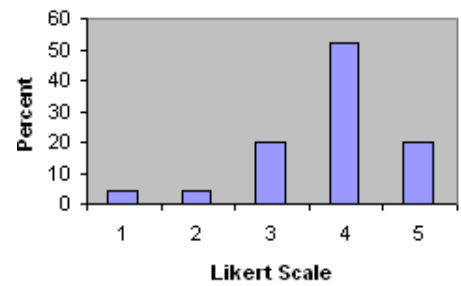

Q. # 7: Do you think that prototype model is a good option to be used in student's projects?

In so far as to this question, 64 percent of the people said that prototype model is a very good option for the student's projects. Similarly there were only 16 percent of the people who disagreed to use prototype model in student's projects. In the survey of this question only 20 percent of the people did not cast their vote.

The outcome of the analysis of this query tells that prototype model is a very good option to be used in student's projects. This is because the selection o this model facilitates the students to get the SW requirements in detail.

The graphical picture of Table 9 is given by Bar chart 7 below.

Table 9 Prototype Model

| | Likert Scale | Frequency | Percent | Cumulative Percent |
|---|---|---|---|---|
| Valid | 1 | 2 | 8.0 | 8.0 |
| | 2 | 2 | 8.0 | 16.0 |
| | 3 | 5 | 20.0 | 36.0 |
| | 4 | 12 | 48.0 | 84.0 |
| | 5 | 4 | 16.0 | 100.0 |
| | Total | 25 | 100.0 | |

Bar chart 7 Prototype Model

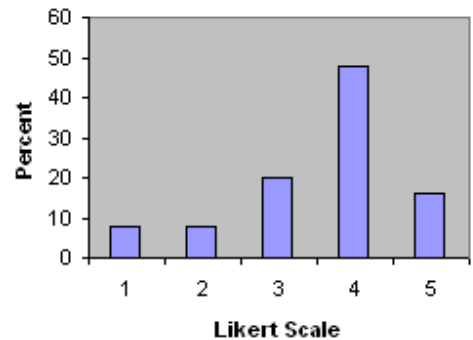

Q. # 8: Is it necessary to provide practical experience and knowledge of the SPM to the students of SW engineering?

According to the results of question 8 given in Table 10, 60 percent of the people had the same opinion to support this fact that it is necessary to teach SPM to the students of SW engineering by focusing on the practical experience and knowledge so that the students will be trained like pilots for fighting in the field of battle (SI). Concomitantly only 24 percent of the people gave their point of view in different to the agreed people. You can see from the Table 10 that only 16 percent people who remained in neutral state.

So the favorable percentage of the Table 10 informs that it is vital to provide practical experience and knowledge of the SPM to the students of SW engineering.

Results are shown in the form of table namely Table 10 and bar chart 8 below.

Table 10 Practical Knowledge of SPM

| Likert Scale | Frequency | Percent | Cumulative Percent |
|---|---|---|---|
| 1 | 3 | 12.0 | 12.0 |
| 2 | 3 | 12.0 | 24.0 |
| 3 | 4 | 16.0 | 40.0 |
| 4 | 12 | 48.0 | 88.0 |
| 5 | 3 | 12.0 | 100.0 |
| Total | 25 | 100.0 | |

Following bar chart expose the results of Table 10.

Bar chart 8 Practical Knowledge of SPM

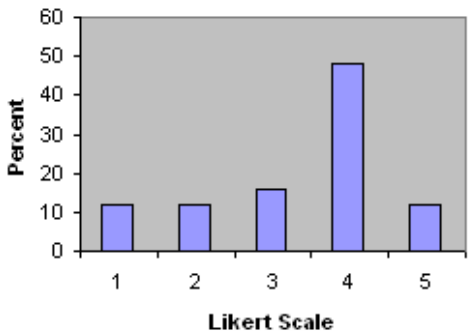

Q. # 9: Do you think that the proposed solution will enable students to get sound knowledge of the subject (SPM)?

The outcomes of survey mentioned in Table 11 reveals that 68 percent of the people had the same conclusion that the proposed solution of this paper will definitely enable the students to get sound knowledge of the SPM. In conjunction with, the 12 percent of the people did not agree with 68 percent that the proposed solution will not enable students to get sound knowledge of the subject. Resultantly opposing factor is too much less than the supporting one. This clearly highlighting that the proposed solution will be helpful to the students to acquire sound knowledge of SPM. Just 20 percent of the people were in the neutral condition.

Hence, this can easily be derived from the survey results of this query in question that the proposed solution will facilitate the students to get sound knowledge of the subject

Table 11 Proposed Solution for Sound Knowledge of SPM

| Likert Scale | Frequency | Percent | Cumulative Percent |
|---|---|---|---|
| 2 | 3 | 12.0 | 12.0 |
| 3 | 5 | 20.0 | 32.0 |
| 4 | 10 | 40.0 | 72.0 |
| 5 | 7 | 28.0 | 100.0 |
| Total | 25 | 100.0 | |

Bar chart 9 shows the results in easy style.

Bar chart 9 Proposed Solution for Sound Knowledge of SPM

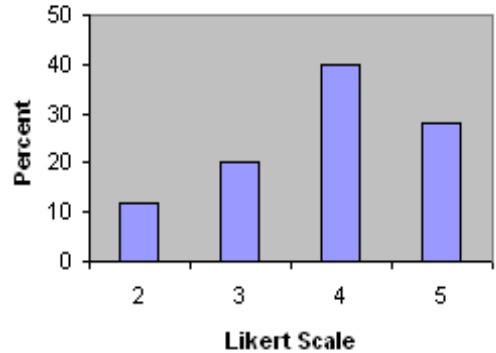

Q. # 10: Do you think that the proposed solution provides the students with hands-on practice of SPM?

As mentioned in Table 12, 88 percent of the people had same eye to eye on the same point that the proposed solution will have to impart hands-on practice of SPM for the students of SW engineering where as contrasting aspect is 8 percent that is very minor difference as compared to the supporting percentage. This compare & contrast undoubtedly underline the truth that the proposed solution will give hands-on practice of the SPM to the students of SW engineering. Point to be noted that only 4 percent of the people (just one frequency) had neutral status. This give the clear cut strong argument that almost all of the people who surveyed expressed their keen interest in this question.

Therefore this further leads to strongly conclude that the authors proposed solution absolutely assure that the said solution certainly gives hands-on practice of SPM to the future SW engineers.

Table 12 Proposed Solution for Hands-on Practice

| Likert Scale | Frequency | Percent | Cumulative Percent |
|---|---|---|---|
| 2 | 2 | 8.0 | 8.0 |
| 3 | 1 | 4.0 | 12.0 |
| 4 | 15 | 60.0 | 72.0 |
| 5 | 7 | 28.0 | 100.0 |
| Total | 25 | 100.0 | |

The 4th bar in Bar chart 10 shows the trustworthy percentage with the mega difference under the query in question.

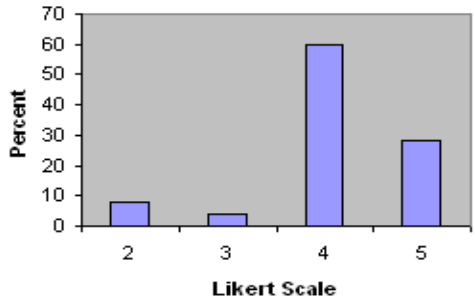

Bar chart 10 Proposed Solution for Hands-on Practice

Q. # 11: Does assigning different roles to the students in SW projects provide them with better understanding of the SPM?

Consistent with the upshot of the question 11 displayed in Table 13, 64 percent of the people were arguing that assigning different roles to the students in SW projects give them better understanding of the SPM. This is because of the verity that there are different roles that to be played during the development of the SW project. As far as the opposing percent concerned 16 percent of the people gave their favor in opposite in question while 20 percent of the people stay neutral.

So when the SW engineering students learnt during their study that how to act at what role/phase of the SDLC (SW Development Life Cycle) then the students will feel confidence to handle different types of problems/challenges in practical life of SI.

Bar chart 11 shows the tabular data of this question in chart form.

Table 13 Better Understanding of SPM

| | Likert Scale | Frequency | Percent | Cumulative Percent |
|---|---|---|---|---|
| Valid | 2 | 4 | 16.0 | 16.0 |
| | 3 | 5 | 20.0 | 36.0 |
| | 4 | 14 | 56.0 | 92.0 |
| | 5 | 2 | 8.0 | 100.0 |
| | Total | 25 | 100.0 | |

Bar chart 11-Better Understanding of SPM

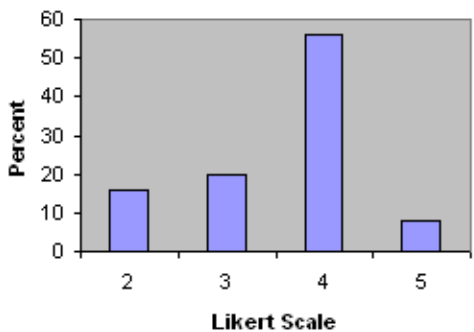

udents with SI to remain in touch with its updates?

Now the authors going to present the culminations of question 12 provided in Table 14, 68 percent of the people stance in the goodwill of this query. For the moment 24 percent of the people voted in contradictory direction. Despite the fact that only 8 percent of the people were hang Here is the bar chart of the above Table.

Bar chart 12 Effectiveness of Interaction with SI

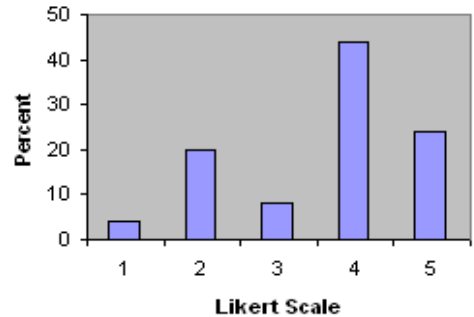

around in neutral mode.

This supportive and favorable percentage, mentioned in Table 14, explains that the interaction of the students of the SPM with SI is very effective and productive.

Table 14 Effectiveness of Interaction with SI

| | Likert Scale | Frequency | Percent | Cumulative Percent |
|---|---|---|---|---|
| Valid | 1 | 1 | 4.0 | 4.0 |
| | 2 | 5 | 20.0 | 24.0 |
| | 3 | 2 | 8.0 | 32.0 |
| | 4 | 11 | 44.0 | 76.0 |
| | 5 | 6 | 24.0 | 100.0 |
| | Total | 25 | 100.0 | |

Q. # 13: Do you think that the proposed solution provides better method to understand the cost estimation of the SW project?

The query in question is very interesting because it is about the SW cost estimation. The summary of the survey of this question is cited in the following table. Domino effect of this table describes that 48 percent of the people were in encouraging direction. In the intervening time merely 16 percent of the people were discouraging this query.

It is notable to see that 36 percent of the people linger neutral.

Thus supportive percentage tells that the proposed solution is very much helpful for the students of the SW engineering to find out the estimated SW cost because it deeply analyze the system to obtained maximum number of SW requirements.

Table 15 Proposed Solution for Better SW Cost Estimation

| | Likert Scale | Frequency | Percent | Cumulative Percent |
|---|---|---|---|---|
| Valid | 2 | 4 | 16.0 | 16.0 |
| | 3 | 9 | 36.0 | 52.0 |
| | 4 | 9 | 36.0 | 88.0 |
| | 5 | 3 | 12.0 | 100.0 |
| | Total | 25 | 100.0 | |

Here is the bar chart of the above Table.

Bar chart 13 Proposed Solution for Better SW Cost Estimation

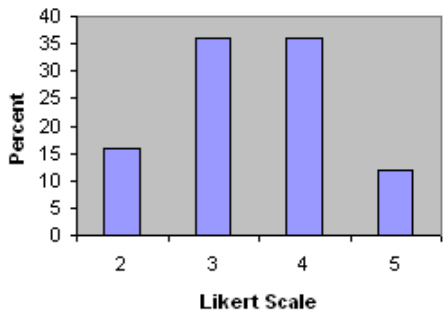

Q. # 14: Do you think the proposed solution is faster as compared to other solutions for better understanding of the SPM?

Table 16 illustrates the outcome of the query in question. As much as the evaluation of this question concerned, 64 percent of the people made their choice in the support of the proposed solution. Temporarily, 16 percent of the people were not supportive. This is a small percentage as compared to the supportive one. It means that the proposed solution of this research is really a faster method of understanding of the SPM for the students of SW engineering. Simultaneously 20 percent of the people had neutral opinion.

In comparison of the favorable percentage with unfavorable one, it strongly means that the majority of the people were accepting the proposed solution of this research as a very fast and quick method for the better understanding of the SPM to train the future SW engineers.

Table 16 Comparison of Existing & Proposed Solution for the Understanding of SPM

| Likert Scale | Frequency | Percent | Cumulative Percent |
|---|---|---|---|
| Valid 1 | 1 | 4.0 | 4.0 |
| 2 | 3 | 12.0 | 16.0 |
| 3 | 5 | 20.0 | 36.0 |
| 4 | 11 | 44.0 | 80.0 |
| 5 | 5 | 20.0 | 100.0 |
| Total | 25 | 100.0 | |

Bar chart 14 Comparison of Existing & Proposed Solution for the Understanding of SPM

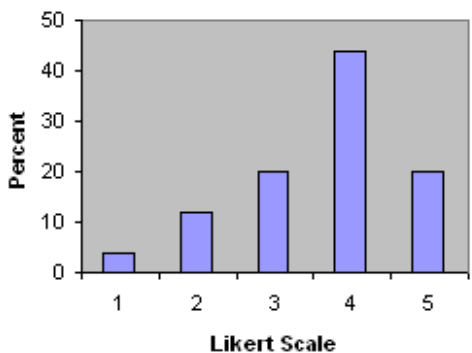

The questionnaire for the survey consists of 14 questions that were disturbed among different Pakistani Universities. The authors received 25 responses for all the questions of this research, which were analyzed using SPSS (statistical analyzing SW). Cumulative result of 14 questions of this research is given in Table 17. As far as the evaluation of the proposed solution of this research concerned, 64.3 percent of the people were supporting this research. It means that the authors have done this research very nicely and accurately. It is a clear cut picture regarding the people who were opposing the proposed solution, which is just 14.8 percent. It further clarified that the percentage of the opposing people is less than quarter time of the percentage of the supportive people. But 21 percent of the people responded in neutral. Consequently, the compared and contrasted outcomes of this question truly display the successful picture of this research. This proposed teaching model will defiantly contributed to reduce the failure rate of the SW project in SI in one hand and to teach the SPM practically in other hand.

Table 17 Cumulative Result

| Likert Scale | Frequency | Percent | Cumulative Percent |
|---|---|---|---|
| Valid 1 | 12 | 3.0 | 3.0 |
| 2 | 47 | 11.8 | 14.8 |
| 3 | 84 | 21.0 | 35.8 |
| 4 | 155 | 38.8 | 74.5 |
| 5 | 102 | 25.5 | 100.0 |
| Total | 400 | 100.0 | |

The bar cart 15 is showing the results in percent mentioned in Table 17.

Bar chart 15 Cumulative Result

Next follows the conclusion to the cumulative result and the research undertaken in this paper.

## 6. Conclusion and Future Work

The author's proposed solution is useful to teach SPM to the students of SW engineering in a better way. This solution is a dowry and provides a practical oriented platform to the students of SPM. The proposed model of teaching SPM also sharpens the following skills of the students, that how to:

1. Gather system requirements
2. SW cost estimation
3. Perform different types of roles (e.g. Project Manager, Designer, Developer, Quality Assurance Engineer etc) in SI.
4. Interaction with the clients (SI)
5. Keep in touch with day to day updates of information technology.
6. Handle the challenges of SW failure.
7. Satisfy the client with sound argument
8. Struggle in the competition of bidding a mega project

Limitations of the related work are handled in this paper to our own extent. There is not a single paper which gave the importance to SI collaboration and hands on-practice for SW engineering students. If the proposed SW project model of this research is adopted in Pakistani Universities then this will lead to the reduction of the failure rate of SW projects.

This will further strengthen the SW houses in terms of output & efficiency by decreasing failure rate.

In future the authors intend to implement the proposed model into the academic environment. The evaluation of the success of the model will be done as future work.